\documentclass[prb,amsmath,showpacs,draft,amsmath,amsfonts,twocolumn,superscriptaddress]{revtex4}

\usepackage{bm}
\begin{document}

\title{Magnetoelasticity theory of incompressible quantum Hall liquids}

\author{I.~V.~Tokatly}

\email{ilya.tokatly@physik.uni-erlangen.de}

\affiliation{Lerhrstuhl f\"ur Theoretische Festk\"orperphysik,
  Universit\"at Erlangen-N\"urnberg, Staudtstrasse 7/B2, 91058
  Erlangen, Germany}

\affiliation{Moscow Institute of Electronic Technology,
 Zelenograd, 124498 Russia}
\date{\today}

\begin{abstract}
A simple and physically transparent magnetoelasticity
theory is proposed to describe linear dynamics of incompressible fractional
quantum Hall states. The theory manifestly satisfies the Kohn theorem and
the $f$-sum rule, and predicts a gaped intra-Landau level collective mode
with a roton minimum. In the limit of vanishing bare mass $m$
the correct form of the static structure factor, $s(q)\sim
q^4$, is recovered. We establish a connection of the present
approach to the fermionic Chern-Simons theory, and discuss further
extensions and applications. We also make an interesting analogy of
the present theory to the theory of visco-elastic fluids.
\end{abstract}


\maketitle 

\section{Introduction} 

The fractional quantum Hall (FQH) effect, which
occurs in a two-dimensional (2D) electron gas subjected to a
strong magnetic field, is one of the most interesting macroscopic
manifestations of quantum mechanics
\cite{QHEbyChakraborty,QHEbyYoshioka}. It is well established that the
perfect quantization of the Hall conductance at particular rational
filling factors $\nu$ is related to the formation of new strongly correlated
states of matter -- incompressible quantum liquids. Despite a substantial
progress in understanding of the FQH effect (mainly based on a
construction of trial wave functions
\cite{Laughlin1983,Jain1989a,Jain1990,CompositeFermions}
or on approximate calculations within Chern-Simons (CS) field theories
\cite{LopFra1991,LopFra1993,SimHal1993,MurSha2003,CompositeFermions})
a simple phenomenological theory of dynamics of FQH states is still
lacking. 

One of the key phenomenological features of incompressible FQH liquids
is a highly collective response to external perturbations. In this
respect FQH liquids look surprisingly similar to the ``classical''
ideal liquids and solids, despite microscopic mechanisms behind the
collective behavior are very different. The collective response of
classical condensed matter leads to a great simplification and
unification of the long wavelength theory, which takes a universal
form of the classical continuum mechanics
\cite{LandauVI:e,LandauVII:e}.  In the present paper we exploit the
abovementioned similarity, and demonstrate that the formalism of
continuum mechanics can be extended to describe collective
dynamics in the FQH regime.  Using general conservation laws, symmetry
and sum-rules arguments we derive a magnetoelasticity theory of
incompressible FQH fluids. For the sake of clarity in this paper we
present a ``minimal'' construction, which is, in a certain sense,
analogous to a single mode approximation (SMA) by Girvin, MacDonald,
and Platzman \cite{GirMacPla1986}. The minimal theory is
parametrized by three elastic constants: the bulk modulus $K$, the
{\it high-frequency} shear modulus $\mu^{\infty}$ (the static shear
modulus of any liquid vanishes), and a new ``magnetic'' modulus
$\Delta$ that is related to the intra-Landau level (LL) excitation gap
\cite{note1a}. Physically the new elastic modulus is responsible for a
``Lorentz shear stress'' induced by time-dependent shear deformations
in a system with broken time-reversal symmetry.  We also show that
this three-parameter continuum mechanics can be derived from the
fermionic CS theory within the random phase approximation (RPA)
\cite{LopFra1991,LopFra1993,SimHal1993,CompositeFermions}. However,
the magneto-elastic theory goes beyond presently available field
theoretical approximations. In particular, in its most general form
\cite{note1b}, it predicts two gapped collective modes, which is in
agreement with recent experimental observations
\cite{Hirjibehedin2005}.

The structure of the paper is the following. In Sec.~II we present a
simple phenomenological derivation of the basic equations of linear
magnetoelasticity theory. We calculate linear response functions and
analyze the dispersion relation of collective excitations. In Sec.~III
we demonstrate that our minimal magnetoelasticity is directly
related to the fermionic CS theory when the later is treated at the
level of the time-dependent mean field approximation (RPA). The
conclusion, and the summary of main results are presented in
Sec.~IV. In this section we also discuss an analogy of the present
magnetoelasticity theory to the hydrodynamics of highly viscous
fluids.

\section{Phenomenological derivation of magnetoelasticity theory}

Let us consider a 2D many-particle system in the presence of a strong
magnetic field ${\bf B}$ perpendicular to the plane. For definiteness
we assume that the particles are confined to the $x,y$ plane, while
the magnetic field is pointing in the $z$-direction, ${\bf B}= {\bf
e}_{z}B$. The dynamics of the system should satisfy the following
exact local conservation laws for the number of particles and momentum
\begin{eqnarray} 
(\partial_{t} + {\bf v}\nabla)n &+& n\partial_{\alpha}v_{\alpha} = 0,
\label{1}\\
m(\partial_{t} + {\bf v}\nabla)v_{\alpha} &+& B\varepsilon_{\alpha\beta}v_{\beta} 
+ \frac{1}{n}\partial_{\beta}P_{\alpha\beta} + \partial_{\alpha}U_{\text{H}}
= F_{\alpha} 
\label{2}
\end{eqnarray}
where $n$ and ${\bf v}={\bf j}/n$ are the density and the velocity of
the quantum fluid, $U_{\text{H}}$ is the Hartree potential, and ${\bf
F} = -\nabla U + \partial_{t}{\bf a}$ is the force due to external
time-dependent scalar and vector potentials \cite{units}.  According
to mapping theorems of the time-dependent density functional theory
(TDDFT) \cite{RunGro1984,Vignale2004}, the exact stress tensor
$P_{\mu\nu}$ is a unique functional of the velocity ${\bf v}$ (for a
detailed discussion of a hydrodynamic formulation of TDDFT based on
the local conservation laws of Eqs.~(\ref{1}) and (\ref{2}) see
Ref.~\onlinecite{TokPRB2005a,TokPRB2005b} and references therein).
For general nonlinear dynamics the functional $P_{\mu\nu}[{\bf v}]$
is unknown. However, in the linear regime its structure can be easily
established from the symmetry considerations.  Let us introduce the
displacement vector ${\bf u}({\bf r},t)$, which is related to the
velocity ${\bf v}({\bf r},t)$, and to the density $n({\bf
r},t)=n_{0}+\delta n({\bf r},t)$, as follows: ${\bf
v}=\partial_{t}{\bf u}$, and $\delta n = -n_{0}\nabla{\bf
u}$. Linearizing Eqs.~(\ref{1}) and (\ref{2}) we arrive at the
equation of the exact linear continuum mechanics
\begin{equation} 
-m\omega^{2}u_{\alpha} -i\omega
B\varepsilon_{\alpha\beta}u_{\beta} +
\frac{1}{n_0}\partial_{\beta}\delta P_{\alpha\beta} +
\partial_{\alpha}U_{\text{H}} = F_{\alpha},
\label{3}
\end{equation} 
where $\delta P_{\alpha\beta}[{\bf u}]$ is a linear functional of
${\bf u}$. As usual, we decompose the stress tensor $\delta
P_{\alpha\beta}$ into a scalar (pressure) and traceless (shear)
parts
\begin{equation} 
\delta P_{\alpha\beta} =
\delta_{\alpha\beta}\delta P + \pi_{\alpha\beta}, 
\qquad {\rm Tr}\hat{\bf \pi}=0. 
\label{4a}
\end{equation} 
Rotational symmetry in the $x,y$-plane, and the
presence of an axial vector perpendicular to the plane dictate the
following most general form for linear functionals $\delta P[{\bf
  u}]$ and $\pi_{\alpha\beta}[{\bf u}]$ (in fact, for their
Fourier components)
\begin{eqnarray} 
\delta P &=& - K u_{\gamma\gamma},
\label{4}\\ 
\pi_{\alpha\beta} &=& - {\mu}(2u_{\alpha\beta}-
\delta_{\alpha\beta}u_{\gamma\gamma})
+i\omega{\Lambda}(\varepsilon_{\alpha\gamma}u_{\gamma\beta}
+\varepsilon_{\beta\gamma}u_{\gamma\alpha})
\label{5}
\end{eqnarray} 
were $u_{\alpha\beta} = (\partial_{\alpha}u_{\beta}+
\partial_{\beta}u_{\alpha})/2$ is the strain tensor. In general the
coefficients $K$, ${\mu}$ and ${\Lambda}$ in Eqs.~(\ref{4}), (\ref{5})
are allowed to be functions of $\omega$ and $|{\bf q}|=q$. The scalar
$\delta P$, Eq.~(\ref{4}), defines the trace of the stress tensor
$\delta P_{\alpha\beta}$, which is the change of pressure due to
density variations. The first term on the right hand side of
Eq.~(\ref{5}) is easily recognized as the usual shear stress induced
by the shear strain \cite{LandauVII:e}. Hence the coefficients $K$ and
$\mu$ correspond to the exact ($\omega$- and $q$-dependent) bulk and
shear moduli respectively. To reveal the physics of the second term in
Eq.~(\ref{5}), we note that the expression in brackets in this term is
simply a symmetrized cross-product of ${\bf e}_{z}$ and the strain
tensor $u_{\alpha\beta}$. Apparently only the shear strain tensor,
$S_{\alpha\beta} = 2u_{\alpha\beta}-
\delta_{\alpha\beta}u_{\gamma\gamma}$, contributes to that
cross-product. Hence Eq.~(\ref{5}) can be rewritten in the following
simple structural form
\begin{equation} 
\hat{\bm\pi} = - \mu\hat{\bf S} -
{\Lambda}\partial_{t}\hat{\bf S}\times{\bf e}_{z},
\label{6}
\end{equation} 
The representation of Eq.~(\ref{6}) suggests a natural physical
interpretation of the second contribution to the shear stress
tensor. This is a local ``Lorentz stress'', which is proportional to
the rate of shear strain. The divergence of the Lorentz stress tensor
gives a transverse compression force exerted on two neighboring
fluid layers moving in opposite directions.

The Kohn theorem and the $f$-sum rule bound the possible $\omega$- and
$q$-dependence of $K$, $\mu$ and $\Lambda$. The Kohn theorem requires
the stress force, $\partial_{\beta}\delta P_{\alpha\beta}$, in
Eq.~(\ref{3}) to vanish for a rigid motion. This requirement forbids
divergencies of the elastic moduli in the limit $q\to 0$. The $f$-sum
rule is equivalent to the statement that at 
$\omega\to\infty$ the leading contribution to the left hand side of
Eq.~(\ref{3}) is given by the first (acceleration) term. Thus the Kohn
theorem as well as the $f$-sum rule are satisfied if the functions
$K(\omega,q)$, $\mu(\omega,q)$, and $\Lambda(\omega,q)$ have finite
$q\to 0$ and $\omega\to\infty$ limits. Indeed, in this case the stress
force $\partial_{\beta}\delta P_{\alpha\beta}$ is of the order of
$q^{2}$ in the small-$q$ limit, so that at $q\to 0$ the equation of motion,
Eq.~(\ref{3}), simplifies as follows
\begin{equation}
-m\omega^{2}{\bf u} -i\omega{\bf u}\times{\bf B} 
+ {\bf q}({\bf q u})n_{0}V(q) + O(q^{2}) = {\bf F}.
\label{3a} 
\end{equation}
The third term on the left hand side in Eq.~(\ref{3a}) corresponds to
the Hartree contribution [$V(q)$ is the interaction potential], while
$O(q^{2})$ represents the stress force that is proportional to
$q^{2}$. Apparently a solution to Eq.~(\ref{3a}) satisfies both the Kohn
theorem and the $f$-sum rule, provided the coefficients in the
$O(q^{2})$ term (which are composed of elastic moduli) are finite at
$\omega\to\infty$. For a system with Coulomb interaction, $V(q)=2\pi/q$, the
Hartree term in Eq.~(\ref{3a}) is proportional to $q$, which yields a
known subleading correction $\sim q$ to the dispersion of the Kohn
mode. Finally, one more obvious restriction on $K$, $\mu$, and $\Lambda$
is that in the limit of a strong magnetic field ($m\to 0$) all elastic
moduli should contain only the interaction energy scale.

Up to this point the discussion was absolutely general. Now we
concentrate on a particular case of incompressible FQH liquids.
Let us substitute the general form of the stress tensor,
Eqs.~(\ref{4}), (\ref{5}), into Eq.~(\ref{3}). Considering $m\to 0$
limit (i.~e. neglecting the acceleration term) we get the
following equation for the displacement vector
\begin{equation} 
-i\omega (n_{0}B + {\Lambda}q^{2}){\bf
u}\times{\bf e}_{z} + {\mu}q^{2}{\bf u} + K_{q}{\bf q}
({\bf qu}) = n_{0}{\bf F},
\label{7}
\end{equation} 
where $K_{q}=K+n_{0}^{2}V(q)$. Equation (\ref{7}) describes the lowest
LL dynamics. By setting ${\bf F}=0$, we obtain a linear homogeneous
equation that determines intra-LL eigenmodes. Solution of this
equation leads to the following, formally exact dispersion equation for
allowed frequencies of eigenmodes
\begin{equation} 
\omega^{2} = \frac{{\mu}({\mu} +
K_{q})}{(n_{0} + {\Lambda}q^{2}l^{2})^{2}}q^{4}l^{4},
\label{8}
\end{equation} 
The most important feature of incompressible FQH liquids is the gap in
the spectrum of intra-LL excitations. This fact can be used to
establish a frequency dependence of the shear modulus $\mu$. The exact
dispersion equation of Eq.~(\ref{8}) shows that a finite gap $\Delta$
at $q\to 0$ can exist only if the elastic moduli diverge either in the
$q$ domain (as $1/q^4$) or in the $\omega$ domain (when $\omega\to
\Delta$). However, the singularities in $q$ are forbidden by the Kohn
theorem. Hence, the only allowed behavior for the exact elastic moduli
is to diverge at $\omega\to \Delta$. The most natural assumption would
be $\mu\sim (\omega^2 - \Delta^2)^{-1}$. This expectation
can be rigorously confirmed by the following arguments. The exact
density response function $\chi(\omega,q)$, which straightforwardly
follows from the relation $\delta n = - n_{0}\nabla{\bf u}$, and
Eq.~(\ref{7}) with ${\bf F}=-\nabla U$, takes the form \cite{note4}
\begin{equation} 
\chi(\omega,q) = \frac{n_{0}^{2}{\mu}}{\omega^2(n_{0}
+ {\Lambda}q^{2}l^{2})^{2}
- {\mu}({\mu} + K_{q})q^{4}l^{4}}q^{4}l^{4}
\label{9}
\end{equation} 
In the $q\to 0$ limit, Eq.~(\ref{9}) reduces to $\chi =
{\mu}q^{4}l^{4}/\omega^2$. On the other hand, in that limit one
expects \cite{GirMacPla1986} to have only one gapped intra-LL mode and the
response function of the form $\chi\sim
(ql)^4/(\omega^2-\Delta^2)$. The above two representations of the
density response function uniquely determine the small-$q$ form of the
exact dynamic shear modulus ${\mu}(\omega)$:
\begin{equation} 
{\mu}(\omega) = \frac{\omega^2}{\omega^2 - \Delta^2}\mu^{\infty},
\label{10}
\end{equation} 
where $\mu^{\infty}$ is a constant that has a clear meaning of the
high-frequency shear modulus. In the low frequency limit the effective
shear modulus $\mu$, Eq.~(\ref{10}), vanishes as it should
do in a liquid state. 

A resonant frequency dependence of the form
Eq.~(\ref{10}) commonly appears in precession dynamics. Therefore
it is natural to assume that the shear stress tensor
$\pi_{\alpha\beta}$ experiences a torque
$\Delta(\hat{\bm\pi}\times{\bf e}_{z})$, and to guess the following
equation of motion for $\pi_{\alpha\beta}$
\begin{equation}
\partial_{t}\hat{\bm\pi} + \Delta(\hat{\bm\pi}\times{\bf e}_{z}) = 
-\mu^{\infty}\partial_{t}\hat{\bf S}.
\label{11}
\end{equation} 
The cross product of a symmetric second rank tensor and the unit vector
${\bf e}_{z}$ is defined after Eqs.~(\ref{5}) and (\ref{6}), namely
$$
(\hat{\bm\pi}\times{\bf e}_{z})_{\alpha\beta} = 
  (\varepsilon_{\alpha\gamma}\pi_{\gamma\beta} + 
   \varepsilon_{\beta\gamma}\pi_{\gamma\alpha})/2.
$$
By solving Eq.~(\ref{11}), we indeed obtain tensor $\pi_{\alpha\beta}$
of the required general form, Eq.~(\ref{5}), i.~e.
$$ 
\hat{\bm\pi} = - \mu(\omega)\hat{\bf S} +
i\omega\Lambda(\omega)\hat{\bf S}\times{\bf e}_{z},
$$
with $\mu(\omega)$ of Eq.~(\ref{10}), and $\Lambda(\omega)$ defined by
the following expression
\begin{equation} 
\Lambda(\omega) =
\frac{\mu^{\infty}\Delta}{\omega^2 - \Delta^2}
\label{12}
\end{equation} 
Thus, assuming only the presence of the gap in the spectrum, and
making a plausible guess about an equation of motion for the shear
stress tensor, we were able to recover the frequency dependence of
both the shear modulus $\mu$, and the magnetic modulus $\Lambda$. At
this point it is worth noting an interesting analogy of the present
derivation and the theory of visco-elastic liquids. In fact, the line
of arguments, which led us from Eq.~(\ref{10}) to Eq.~(\ref{11}), is
very similar to the heuristic derivation of the Maxwellian theory of
highly viscous fluids \cite{LandauVII:e} (for a more detailed
discussion see Sec.~IV).

In the limit $\omega\to 0$ the magnetic modulus $\Lambda(\omega)$
approaches a constant, while the shear modulus $\mu(\omega)$ vanishes
as $\omega^{2}$, which is quite remarkable. Solving Eq.~(\ref{7}) with
proper external fields we find that exactly this low-frequency
behavior, $\mu\sim\omega^{2}$, is required to guarantee the correct
low-energy response of FQH liquid, i.~e., the proper static Hall
conductivity and a creation of a localized fractional charge by the
adiabatic insertion of a magnetic flux.

Substituting $\mu(\omega)$, Eq.~(\ref{10}), and
$\Lambda(\omega)$, Eq.~(\ref{12}), in the dispersion equation
of Eq.~(\ref{8}), and assuming a constant bulk modulus $K$,
we find that in the present theory there exists only one collective
mode with the following dispersion 
\begin{equation} 
\omega_{q}^2 = \Delta^2 - 2\Delta\bar{\mu}^{\infty}q^{2}l^{2} +
\bar{\mu}^{\infty}(\bar{\mu}^{\infty} + \bar{K} + n_{0}V(q))q^{4}l^{4},
\label{13}
\end{equation} 
where $\bar{\mu}^{\infty}=\mu^{\infty}/n_{0}$ and $\bar{K}=K/n_{0}$ are
the high frequency shear modulus and the bulk modulus per particle
respectively. The density response function $\chi(\omega,q)$ and the
static structure factor $s(q)$ take the form
\begin{equation} 
\chi(\omega,q) = \frac{\mu^{\infty}}{\omega^2 -
\omega_{q}^2}q^{4}l^{4}, \qquad 
s(q) = \frac{\mu^{\infty}}{2\omega_{q}}q^{4}l^{4}
\label{14}
\end{equation} 
The collective mode, Eq.~(\ref{13}), is by construction
gapped. A much more surprising feature of Eq.~(\ref{13}) is a negative
curvature at small $q$, and a well defined roton-like minimum at
finite $q$, which is in excellent agreement with the phenomenology of
incompressible FQH liquids.

The final magnetoelasticity theory of incompressible FQH liquids
corresponds to the equation of motion, Eq.~(\ref{3}), with $\delta
P_{\alpha\beta} = -\delta_{\alpha\beta}K(\nabla{\bf u}) +
\pi_{\alpha\beta}$, where $\pi_{\alpha\beta}$ is the solution to
Eq.~(\ref{11}). Thus the closed set of equations of the theory takes
the form
\begin{eqnarray}
m\partial_{t}^{2}{\bf u} &+& 
\partial_{t}{\bf u}\times{\bf B} - \frac{K}{n_{0}}\nabla(\nabla{\bf u}) 
+ \frac{\nabla\hat{\bm\pi}}{n_{0}} + \nabla U_{\rm H} = {\bf F}
\label{final1}\\
\partial_{t}\hat{\bm\pi} &+& \Delta(\hat{\bm\pi}\times{\bf e}_{z}) 
+ \mu^{\infty}\partial_{t}\hat{\bf S} = 0.
\label{final2}
\end{eqnarray}
In contrast to the usual continuum mechanics, the shear stress tensor
enters the theory as one more dynamic variable.  The theory is
parametrized by three "elastic" constants, $K$, $\mu$, and
$\Delta$. It satisfies the Kohn theorem and the $f$-sum rule,
reproduces the correct static response (including the creation of
Laughlin quasiparticles), and predicts the intra-LL mode with a roton
minimum. This theory is the main result of the present paper.

Let us discuss Eqs.~(\ref{final1}) and (\ref{final2}) in some more
detail. Setting $\Delta=0$ in Eq.~(\ref{final2}) we recover the
standard local-in-time shear stress--strain relation, 
$\hat{\bm\pi}=-\mu^{\infty}\hat{\bf S}$. Hence in this limiting case
our theory reduces to the usual continuum mechanics of a
magnetized elastic medium. It can describe, for example, a long
wavelength response of a hexagonal crystal. The collective modes in
this regime are gapless magnetophonons. The properties of the system
with a nonzero modulus $\Delta$ are completely different. In fact, the
theory with $\Delta\ne 0$ reproduces most of known
phenomenological features of FQH liquids. The source of that nice
behavior is a nontrivial ``precession'' dynamics of the shear stress
tensor $\hat{\bm\pi}$,  governed by Eq.~(\ref{final2}). It is worth
mentioning that according to Eq.~(\ref{final2}) the system does not
respond to a static shear deformation [$\mu(\omega=0)=0$], which is
an unambiguous signature of a liquid state of matter. 

The magnetoelasticity theory should become exact for small
wave vectors. The small-$q$ form of Eq.~(\ref{13}), 
$$
\omega_{q} \approx \Delta - \bar\mu^{\infty} q^{2}l^{2},
$$ 
shows that the $q\to 0$ curvature of the dispersion of the intra-LL
collective mode is determined solely by the high-frequency shear
modulus $\mu^{\infty}$. This fact opens up a possibility to access
experimentally the shear modulus of FQH liquids.  At small $q$ the
static structure factor, Eq.~(\ref{14}), takes the form
$
s(q)\approx
(\mu^{\infty}/2\Delta)q^{4}l^{4}
$
(which coincides with the result of Ref.~\onlinecite{ConVig1998}).
Using the static structure factor for the Laughlin state 
$
s(q)\approx n_{0}[(1-\nu)/8\nu]q^{4}l^{4}
$
,\cite{GirMacPla1986} we can 
relate $\mu^{\infty}$ to $\Delta$ as follows
$\bar\mu^{\infty}=(1-\nu)\Delta/4\nu$. This 
relation leads to an interesting universal result for the small-$q$
dispersion:
\begin{equation}
\frac{\omega_{q}}{\Delta} \approx 1 - \frac{1-\nu}{4\nu}q^{2}l^{2}
\label{curvature}
\end{equation} 
The normalized dispersion relation of Eq.~(\ref{curvature}) is indeed in
good agreement with the results of numerical calculations within
SMA for $\nu=1/3,1/5,1/7$. \cite{GirMacPla1986}

\section{Connection to the fermionic Chern-Simons theory}

The main
advantage of any continuum mechanics lies in its universality. It should
provide a long wavelength limit of any microscopic theory that
correctly captures the key physics of the
problem. Below we recover the three-parameter magnetoelasticity
theory from the fermionic CS theory with the time-dependent mean-field
approximation (which is equivalent to RPA) \cite{SimHal1993}.

Consider a system in an incompressible state with the filling factor
$\nu=p/(2sp+1)$. (Within the composite fermion concept, $2s$ is the
number of flux quanta attached to every electron, and $p$ is the
number of completely filled composite fermion LL
\cite{Jain1989a,Jain1990,CompositeFermions}). Our goal is to describe
the long wavelength dynamics in the presence of weak external
potentials, $U({\bf r},t)$ and ${\bf a}({\bf
r},t)$. In the time-dependent mean field approximation 
the fermionic CS theory \cite{CompositeFermions,SimHal1993}
reduces to the following equation for the one particle density
matrix $\rho({\bf r}_{1},{\bf r}_{2},t)$
\begin{equation}
i\partial_{t}\rho = [\hat{H}_{\text{CS}},\rho],
\label{15} 
\end{equation}
where $\hat{H}_{\text{CS}}$ is the mean field CS Hamiltonian:
\begin{equation}
\hat{H}_{\text{CS}} = \frac{1}{2m^{*}}[-i\nabla + 
\bm{\mathcal A}({\bf r},t)]^{2} + U_{\text{eff}}({\bf r},t).
\label{16} 
\end{equation}
Here $m^{*}$ is the effective mass of composite fermions, and
$U_{\text{eff}}=U_{\text{H}}+U$ and $\bm{\mathcal A}= {\bf A} + {\bf
A}^{\text{CS}} + {\bf A}^{(1)} + {\bf a}$ are the full
selfconsistent scalar and vector potentials respectively. The static
part, ${\bf A}$, of the full vector potential is related to the
external constant magnetic field,
$B=\varepsilon_{\alpha\beta}\partial_{\alpha}A_{\beta}$. The potential
${\bf A}^{\text{CS}}$ describes the mean CS field that is produced by
$2s$ flux quanta, attached to every electron:
\begin{equation}
B^{\text{CS}} 
= \varepsilon_{\alpha\beta}\partial_{\alpha}A^{\text{CS}}_{\beta} 
= - 4\pi sn,
\label{17}
\end{equation} 
The second selfconsistent vector potential, 
\begin{equation}
{\bf A}^{(1)}= (m^{*} - m){\bf v},
\label{17a}
\end{equation} 
plays a role of $F_{1}$-interaction in the Landau
Fermi-liquid theory \cite{SimHal1993,CompositeFermions}. It allows one to
introduce an effective mass $m^{*}\ne m$, but still
satisfy the requirements of the Galilean invariance, i.~e. the Kohn
theorem and the $f$-sum rule. 

To derive the long wavelength limit of Eq.~(\ref{15}) we
introduce the following ``gauge invariant'' Wigner function
\begin{equation}
f_{\bf k}({\bf r},t) = 
\int d{\bm\xi} \rho({\bf r}+{\bm\xi}/2,{\bf r}-{\bm\xi}/2,t)
e^{-i{\bm\xi}\left[{\bf k}-\overline{\bm{\mathcal A}}({\bf r},{\bm\xi},t)\right]}
\label{19}
\end{equation}
where
$\overline{\bm{\mathcal A}}({\bf r},{\bm\xi},t) = \int_{-1}^{1}
{\bm{\mathcal A}}({\bf r}+\lambda{\bm\xi}/2,t)d\lambda/2$. The density
$n$, the velocity ${\bf v}$, and the stress tensor $P_{\alpha\beta}$
are related to zeroth, first, and second moments of the Wigner
function:
\begin{eqnarray}
&n& = \sum_{\bf k}f_{\bf k}, \quad
{\bf v} = \frac{1}{n}\sum_{\bf k}\frac{\bf k}{m^{*}}f_{\bf k},
\label{20}\\
P_{\alpha\beta} &=& \frac{1}{m^{*}}\sum_{\bf k}
(k_{\alpha}-m^{*}v_{\alpha})(k_{\beta}-m^{*}v_{\beta})f_{\bf k}
\label{21}
\end{eqnarray}
In the homogeneous equilibrium state the Wigner function of
Eq.~(\ref{19}) takes the form
\begin{equation}
f^{(0)}_{\bf k} = 
2e^{-k^{2}/B^{*}}\sum_{n=0}^{p-1}(-1)^{n}L^{0}_{n}(2k^{2}/B^{*}),
\label{22}
\end{equation}
where $L^{m}_{n}(x)$ are Laguerre polynomials, and
$B^{*}=B+B^{\text{CS}}_{0}$ is the effective magnetic field that acts
on composite fermions ($B^{\text{CS}}_{0} = -4\pi sn_{0}$). Equation
(\ref{22}) provides the initial condition for our dynamic problem. 

Let the characteristic length scale $L$ of the external fields be much
larger then the effective magnetic length $l^{*}=1/\sqrt{B^{*}}$. Using
Eqs.~(\ref{15}) and (\ref{19}) we find that the long wavelength
dynamics of the Wigner function is governed by the following simple
equation of motion
\begin{equation}
\frac{\partial f_{\bf k}}{\partial t}
+ \frac{k_{\alpha}}{m^{*}}\frac{\partial f_{\bf k}}{\partial x_{\alpha}} 
+ \left(\frac{\mathcal B}{m^{*}}\varepsilon_{\alpha\beta}k_{\beta}
- \frac{\partial{\mathcal A}_{\alpha}}{\partial t} 
+ \frac{\partial U_{\text{eff}}}{\partial x_{\alpha}}\right)
\frac{\partial f_{\bf k}}{\partial k_{\alpha}} = 0
\label{23}
\end{equation}
where ${\mathcal B} =
\varepsilon_{\alpha\beta}\partial_{\alpha}{\mathcal A}_{\beta}$. 
Despite Eq.~(\ref{23}) is exactly of the form 
of semiclassical Boltzmann equation, the function $f_{\bf k}$ is the
full Wigner function, Eq.~(\ref{19}), of the quantum
system. Formally the quantum mechanics enters the problem via the
initial condition, Eq.~(\ref{22}), for the equation of motion,
Eq.~(\ref{23}). The next step is to linearize Eq.~(\ref{23}) about the
equilibrium solution, Eq.~(\ref{22}), and to derive linearized
equations of motion for the moments, Eqs.~(\ref{20}), (\ref{21}). All
calculations closely follow the derivation of generalized
hydrodynamics in Refs.\onlinecite{TokPPRB1999,TokPPRB2000}. The zeroth
moment of the linearized Eq.~(\ref{23}) gives the continuity
equation
\begin{equation} 
\partial_{t}\delta n + n_{0}\partial_{\alpha}v_{\alpha} = 0,
\label{24}
\end{equation}
Similarly taking the first moment of Eq.~(\ref{23}) we get the force
balance equation of the following form
\begin{eqnarray}
m^{*}\partial_{t}v_{\alpha} &+& B^{*}\varepsilon_{\alpha\beta}v_{\beta} 
- \partial_{t}A^{(1)}_{\alpha} - \partial_{t}A^{\rm CS}_{\alpha}
\nonumber\\
&+&\frac{1}{n_{0}}\partial_{\beta}\delta P_{\alpha\beta} + 
  \partial_{\alpha}U_{\text{H}} = 
-\partial_{\alpha}U + \partial_{t}a_{\alpha}. 
\label{25a}
\end{eqnarray} 
Substituting the definition of ${\bf A}^{(1)}$, Eq.~(\ref{17a}), into
Eq.~(\ref{25a}) we find that the combination of the first and the
third terms in the left hand side of Eq.~(\ref{25a}) reduces to the
correct acceleration term, $m\partial_{t}{\bf v}$. In a quite similar
fashion the CS electric field, $-\partial_{t}{\bf A}^{\rm CS}$,
cancels the CS Lorentz force, ${\bf v}\times{\bf B}^{\rm CS}$ [one can
straightforwardly prove that by calculating the time derivative of
Eq.~(\ref{17}), and using the continuity equation]. As a result of
these cancellations the force balance equation takes the standard
form, which is required by the Galilean invariance:
\begin{equation}
m\partial_{t}v_{\alpha} + B\varepsilon_{\alpha\beta}v_{\beta} 
+ \frac{1}{n_{0}}\partial_{\beta}\delta P_{\alpha\beta} + 
  \partial_{\alpha}U_{\text{H}} = F_{\alpha}. 
\label{25}
\end{equation}
Finally, calculating the second moment of the linearized equation (\ref{23}),
we obtain the equation of motion for the stress tensor $\delta
P_{\alpha\beta} = \delta_{\alpha\beta}\delta P +
\pi_{\alpha\beta}$. Decomposition of the scalar and the traceless
parts of this equation leads to the following equations of motion for
the pressure $\delta P$, and for shear stress tensor
$\pi_{\alpha\beta}$ respectively
\begin{eqnarray}
\partial_{t}\delta P &+& 2P_{0}\partial_{\alpha}v_{\alpha} = 0,
\label{26}\\
\partial_{t}\pi_{\alpha\beta} 
&+& (B^{*}/m^{*})(\varepsilon_{\alpha\gamma}\pi_{\gamma\beta} 
+ \varepsilon_{\beta\gamma}\pi_{\gamma\alpha})
\nonumber\\
&+& P_{0}(\partial_{\alpha}v_{\beta} + \partial_{\beta}v_{\alpha} 
- \delta_{\alpha\beta}\partial_{\gamma}v_{\gamma}) = 0,
\label{27}
\end{eqnarray}
where $P_{0}=\sum_{\bf k}k^{2}f^{(0)}_{\bf k}/2m^{*}$ is the
equilibrium (initial) pressure of the composite fermion system. The
third moment of the Wigner function, which is, in general, present in
the equation for the second moment, introduces higher order gradient
corrections that are irrelevant in the long wavelength limit
\cite{TokPPRB2000}.

The system of Eqs.~(\ref{24}), (\ref{25})-(\ref{27}) provides a
long wavelength ``hydrodynamic'' 
representation of the RPA linear response for CS theory
\cite{CompositeFermions,LopFra1991,LopFra1993,SimHal1993}. To
demonstrate an equivalence of the CS hydrodynamics to our
phenomenological magnetoelasticity we express the velocity 
${\bf v}$ in terms of the displacement vector ${\bf u}$: 
\begin{equation}
{\bf v}=\partial_{t}{\bf u}.
\label{v}
\end{equation}
The representation of Eq.~(\ref{v}) allows us to explicitly integrate
the continuity equation, Eq.~(\ref{24}), and the equation for the
pressure, Eq.~(\ref{26})
\begin{equation}
\delta P = - \frac{2P_{0}}{n_{0}}\partial_{\alpha}u_{\alpha}, \qquad
\delta n = - n_{0}\partial_{\alpha}u_{\alpha}.
\label{pressure}
\end{equation}
Substituting Eqs.~(\ref{v}) and (\ref{pressure}) into Eqs.~(\ref{26}) and
(\ref{27}) we find that they become identical to the equations of
magnetoelasticity theory, Eqs.~(\ref{final1}) and (\ref{final2}),
with $K=2P_{0}$, $\mu^{\infty}=P_{0}$ and $\Delta=2B^{*}/m^{*}$.  It is worth
mentioning that the ideal gas relation $K=2\mu^{\infty}$ is a direct
consequence of the lack of correlations in RPA. Thus, despite the
general structure of the magneto-elastic phenomenology is recovered,
the microscopic elastic constants are clearly incorrect.

\section{Conclusion}

We proposed a magnetoelasticity theory of incompressible FQH
liquids. The theory demonstrates that most of fundamental dynamic
properties of FQH liquids (the response to electric and magnetic
fields, the existence of gapped intra-LL collective modes, etc.) can
be described in a surprisingly simple and clear fashion. The whole
linear response is encoded in the local momentum conservation law,
Eq.~(\ref{final1}), supplemented by the equation of motion,
Eq.~(\ref{final2}), for the shear stress tensor $\hat{\bm\pi}$. In
contrast to the standard classical elasticity theory, which is a
theory of one vector field ${\bf u}$, the description of FQH liquids
requires one more dynamic variable -- the traceless tensor field
$\hat{\bm\pi}$. In fact, all unusual dynamic properties of FQH liquids
can be traced back to the equation of motion for the tensor field
$\hat{\bm\pi}$. In this paper we presented a heuristic derivation of
the simplest version of this equation, Eq.~(\ref{final2}), which
yields a minimal, three-parameter magnetoelasticity theory. Using
this theory we were able to reproduce most of the already known results
and to make some new predictions concerning the dispersion of the
intra-LL mode. In particular, we have shown that at small-$q$ the
dispersion is determined by the high-frequency shear modulus
$\mu^{\infty}$, which, in turn, is related the ground state energy
$E_{0}$ (for a 2D Coulomb system with a quenched kinetic energy
$\mu^{\infty}=E_{0}/8$) \cite{TokPRB2005a}.

Importantly, the time-reversal and the rotational symmetry still allow
for a generalization of our key equation of motion for the shear
stress tensor, Eq.~(\ref{final2}). The most general form of the
theory, as well as its extensions for crystalline and liquid
crystalline quantum Hall states, are presented in a separate paper
\cite{note1b}. The most important result of these generalizations is a
prediction of two gapped intra-LL collective modes in FQH liquids. The
present formulation (despite a certain lack of generality) has a great
advantage to be the simplest, but still nontrivial continuum mechanics
of FQH liquids.  In fact, one of our aims in this work is to demonstrate
a principal possibility of interpreting the behavior of highly unusual
FQH liquids using an intuitively clear and physically transparent
language of continuum mechanics.

Another advantage of the minimal construction is its clear connection
to the well understood mean-field CS theory (Sec.~III).  In addition
to that, there is an interesting analogy of the present formulation of
magnetoelasticity theory, Eqs.~(\ref{final1}), (\ref{final2}), and
the theory of visco-elastic fluids by Maxwell \cite{LandauVII:e}. Both
the Maxwellian hydrodynamics and our magnetoelasticity are formulated
in terms of a system of two coupled equations. These are (i) the local
momentum conservation law (the equation of motion for the displacement
${\bf u}$), and (ii) the equation of motion for the shear stress
tensor $\hat{\bm\pi}$. The first equation takes a universal form of
Eq.~(\ref{final1}). It describes a motion of an infinitesimal fluid
element under a combination of the external and internal stress
forces. The dynamics of the stress tensor is, however, specific for
every system. Physically it is related to a relative motion of
particles inside a fluid element, which depends on details of many-body
correlations.  For highly viscous fluids the equation of motion for
$\hat{\bm\pi}$ takes the form \cite{LandauVII:e}
\begin{equation}
\partial_{t}\hat{\bm\pi} + \frac{1}{\tau}\hat{\bm\pi} + 
\mu^{\infty}\partial_{t}\hat{\bf S} = 0,
\label{viscoelasticity}
\end{equation} 
while for incompressible FQH liquids the shear stress tensor satisfies
Eq.~(\ref{final2}). Equation (\ref{viscoelasticity}) describes the
relaxation dynamics ($\tau$ is the relaxation time), which leads to a
nonzero shear viscosity, and results in a usual overdamped
hydrodynamic mode. In contrast to that, the dynamics of the stress
tensor in FQH liquids correspond to a dissipationless precession [see,
Eq.~(\ref{final2})]. This immediately opens a gap, making the liquid
incompressible, and produces a collective mode with a roton minimum.

\section*{Acknowledgment} 
I am grateful to G. Vignale
for numerous discussions. A
part of this work was completed during my visit to the University of
Missouri-Columbia, supported by NSF Grant No. DMR-0313681.

% \bibliography{journals,books,QuantHall,bose,kinetics,tddft,hydrodynamics,geometry,eqDFT,mypapers,QHelasticity_note}

\begin{thebibliography}{25}
\bibitem[{uni()}]{units}
\bibinfo{note}{In this paper we use a system of units with $e=c=\hbar=1$.}
\bibitem[{not({\natexlab{c}})}]{note4}
\bibinfo{note}{The exact density response function for an unprojected theory
  (finite mass $m$) can be obtained from Eq.~(\ref{9}) by the following
  replacement, $\mu q^2 \to \mu q^2 - mn_0\omega^2$.}
\end{thebibliography}

\end{document}